\documentclass[preprint,showpacs,preprintnumbers,amsmath,amssymb]{revtex4}

\usepackage{graphicx}
\usepackage{dcolumn}
\usepackage{bm}

\newcommand{\dt}{\ensuremath{\ \mathrm{d}}}
\newcommand{\expct}[1]{\ensuremath{\left\langle #1 \right\rangle}}

\begin{document}

\draft

\title{Quantum fields and ``Big Rip'' expansion singularities}

\author{H\'ector Calder\'on}
\email{calderon@physics.montana.edu}
\author{William A.\ Hiscock}
\email{hiscock@montana.edu}

\affiliation{Department of Physics, Montana State
University, Bozeman, Montana 59717 }

\date{November 28, 2004}


\begin{abstract}
The effects of quantized conformally invariant massless fields on
the evolution of cosmological models containing a ``Big Rip''
future expansion singularity are examined. Quantized scalar,
spinor, and vector fields are found to strengthen the accelerating
expansion of such models as they approach the expansion
singularity.
\end{abstract}
\maketitle

Studies of distant Type Ia supernovae \cite{Perlmutter,Riess}
strongly indicate that the expansion of the universe is actually
accelerating. Combined with the analysis of the WMAP cosmic
microwave background \cite{Spergel}, it appears that the universe
is filled with a ``dark energy'' amounting to about 70\% of the
closure density. This dark energy is characterized by an effective
equation of state parameter, $w = p/\rho$, the ratio of the
pressure to the energy density.

The simplest model for the dark energy is that it is a
cosmological constant, $\Lambda$, for which $w=-1$. Quintessence
models of classical scalar fields, which have attracted much
attention \cite{Turner, Caldwell1}, yield values of $w
> -1$.
Models of the dark energy with $ w < -1$ are dubbed ``phantom
energy''. Phantom energy has the odd property that its energy
density increases as the universe expands; it also violates the
dominant energy condition. A recent analysis {\cite {Riess2}}
finds $ w = -1.02 \pm \stackrel {0.13}{\scriptstyle 0.19}$.
Clearly, the possibility that our universe contains a phantom
energy with $ w < -1$ cannot be ruled out at present.

In addition to the odd material properties of phantom energy,
cosmological models in which phantom energy becomes dominant
undergo divergent expansion, leading to a future spacelike
singularity at a finite cosmic time, termed a ``Big Rip''
\cite{Caldwell2}. The neighborhood of cosmological singularities
is widely recognized as one of the very few locations where
quantum gravitational effects may play an important dynamical
role. The possible existence of a future Big Rip allows us to
consider a case in which quantum gravitational effects can be
examined in a predictive fashion, rather than merely studying the
consistency of quantum gravitational models applied to
observations of the early universe. In this Letter, we begin such
a study, by examining the form of the vacuum stress-energy for
quantized conformally invariant scalar, spinor, and vector fields
in Big Rip spacetimes. We seek to determine whether such fields,
at the semiclassical level, will weaken or strengthen the
development of a Big Rip singularity.

For a spatially flat Friedman-Robertson-Walker (FRW) spacetime
with metric
\begin{equation}
    ds^2 = -dt^2 + a^2(t)[dx^2+dy^2+dz^2]\;,
\label{FRWmetric}
\end{equation}
the Einstein equation for cosmology containing a mass fraction
$\Omega_m$ of ordinary and dark matter, and $1-\Omega_m$ of dark
energy, is \cite{Caldwell2}
\begin{equation}
    \left(\frac{\dot a}{a} \right)^2 = H_0^2\left[{\frac{\Omega_m}
    {a^3}} + (1 -\Omega_m)a^{-3(1+w)}\right]\;\;, \label{EFE}
\end{equation}
where $H_0$ is the Hubble constant. This may be solved implicitly
for the scale factor
\begin{equation}
  \frac{1}{H_0}\int_{0}^{a(t)}\frac{\xi^{1/2}}{\left[\Omega_m + (1 -
  \Omega_m)\xi^{-3w}\right]^{1/2}}\dt{\xi} = t.
  \label{ImpSol}
\end{equation}
where the lower boundary of the integral has been chosen so that
the Big Bang occurred when $t$ was $0$. The fact that the integral
is convergent for any value $w < -1$ with $a \rightarrow \infty$
demonstrates that Big Rip occurs at a finite time for such values
of $w$.

We are interested in the behavior near the Big Rip, where quantum
effects are expected to become significant, and so we can
approximate Eq.(\ref{ImpSol}) by
\begin{equation}
  T - t = \frac{1}{H_0 (1- \Omega_m)^{1/2}} \int^\infty_{a(t)}
 \xi^{(1 + 3 w)/2} \dt{\xi}\;,
\end{equation}
where $T$ is the time of the Big Rip, {\it i.e.}, $a(T)
\rightarrow \infty$. This yields, as an approximate description of
the scale factor near the Big Rip,
\begin{equation}
    a(t) \simeq \left[\left(\frac{3}{2}\right)H_0
    \left(1-\Omega_m\right)^{1/2}\vert 1+w \vert
    \left(T-t\right)\right]^{\frac{2}{3(1+w)}}\;.
\label{scalefactor}
\end{equation}

It is well known that the vacuum stress-energy of quantized
conformally invariant fields in an FRW spacetime depends only on
the trace anomaly and the choice of appropriate vacuum state,
determined by the mapping of the FRW spacetime's Cauchy surface to
the conformally related Minkowski or Rindler space. A good summary
of these points is given by Candelas and Dowker \cite{Candelas}.
Spatially flat FRW spacetimes are conformally related to Minkowski
space, and hence the vacuum stress-energy tensor of a quantized
conformally invariant field is given simply by
\begin{eqnarray}
 \lefteqn{\langle T_{\mu\nu}\rangle  =
        \frac{\alpha}{3}\left(g_{\mu\nu}R^{;\sigma}_{;\sigma}
        -R_{;\mu\nu}+RR_{\mu\nu}-\frac{1}{4}g_{\mu\nu}R^2\right)} \nonumber
        \\ \nonumber \\
        & \hspace{1 cm} \mbox{} +\beta\left(\frac{2}{3}RR_{\mu\nu}-R_{\mu}^{\sigma}R_{\nu\sigma}
        +\frac{1}{2}g_{\mu\nu}R_{\sigma\tau}R^{\sigma\tau}
        -\frac{1}{4}g_{\mu\nu}R^2\right)\;,
\label{qst}
\end{eqnarray}

where $R_{\mu\nu}$ is the Ricci tensor, and $R$ is the Ricci
scalar. The constants $\alpha$ and $\beta$ are related to the
number and spins of the fields present (see table \ref{SpCf}).
\begin{table} \caption{\label{SpCf} Spin Coefficients}
\begin{ruledtabular}
  \begin{tabular}{llll}
    Spin & $\alpha$ & $\beta$ \\
    0 & $\frac{1}{2800\, \pi^2}$ & $\frac{1}{2800\, \pi^2}$ \\
    $\frac{1}{2}$ & $\frac{3}{2800\, \pi^2}$ & $\frac{11}{5600 \pi^2}$ \\
    1 & $\frac{-9}{1400\,\pi^2}$ & $\frac{31}{1400\,\pi^2}$
  \end{tabular}
\end{ruledtabular}
\end{table}

Evaluating $\langle T_{\mu\nu} \rangle$ for the metric of
Eq.(\ref{FRWmetric}) with the scale factor given by
Eq.(\ref{scalefactor}), we find that the vacuum energy densities
of the scalar, spinor and vector fields (\expct{\rho_0},
\expct{\rho_{1/2}}, and \expct{\rho_1} respectively) can be
written in the form
\begin{equation}
  \expct{\rho_a} = \expct{T_{00}}\vert_{{\;\rm spin} = a} =
  \frac{P_a}{19440\,\pi^2(T - t)^4 (1 + w)^4}
\label{rho}
\end{equation}
where $a$ = 0, 1/2, 1, and $P_a$ is a second degree polynomial in
$w$,
\begin{eqnarray}
P_0 &= - 5 + 18 w + 27 w^2 \; \\
P_{1/2} &= -5 + 54 w + 81 w^2 \; \\
P_1 &= 205 - 162 w -243 w^2 \;
\label{Poly}
\end{eqnarray}
Of these three polynomials, $P_0$ and $P_{1/2}$ are strictly
positive for all $ w < -1$; $P_1$ is positive for $ w_0 < w < -1$
and negative for $ w < w_0$, where $w_0 = -\frac{1}{3} -
\frac{2}{27}\sqrt{174} \simeq -1.31$. The sign of the energy
densities, by Eq.(\ref{rho}), agree with the signs of the
polynomials as functions of $w$.

For all three spins, the ratio of the expected value of the
pressure to the expected value of the energy density is a constant
\begin{equation}
\expct{w} = \frac{\expct{p_a}}{\expct{\rho_a}} = 1 + 2 w\; ,
\label{stateeq}
\end{equation}
which is independent of the spin of the field.

In the absence of a complete self-consistent theory of quantum
gravity, calculated expectation values of the vacuum stress-energy
of quantized fields in a classical background may be used to
indicate, in a perturbative sense, in which direction inclusion of
quantum effects will change the classical solution. We accomplish
this by examining the effective change in the total stress-energy
tensor when the quantized vacuum energies are included. The
qualitative effect of such inclusion on the dynamics of an FRW
cosmology -- i.e., whether the evolution towards a Big Rip
singularity is strengthened or weakened -- is simple to discern,
due to the high degree of symmetry of the spacetime metric.

In deSitter spacetime, the expectation value of the stress energy
tensor, \expct{T_{\mu\nu}}, and the cosmological constant term,
$\Lambda\,g_{\mu\nu}$, have the same form and are usually dealt as
one term (the first renormalizing the value of the second).
However, in the FRW cosmology with phantom energy ($w < -1$), the
pressure to energy density ratio of the quantized fields' vacuum
stress-energy, $\langle w \rangle$, is not equal to the background
ratio, $w$. For the combination of the phantom energy plus the
quantized fields' vacuum stress-energy we define an effective
equation of state parameter by
\begin{equation}
  \label{weff}
  w_{\text{eff}} \equiv
  \frac{p_{\text{total}}}{\rho_{\text{total}}} =
  \frac{p_{\rm background} +\langle p \rangle}{\rho_{\rm
  background} +\langle \rho \rangle} \; .
\end{equation}

Using Eq.(\ref{stateeq}), this can be rewritten as
\begin{equation}
w_{\rm eff} = w + (1+w)\frac{\langle \rho \rangle}{\rho + \langle
\rho \rangle} \; . \label {weff2}
\end{equation}
If $w_0 < w < -1$, then $\langle \rho \rangle > 0 $ for all of the
conformally invariant quantized fields, which ensures that
\begin{equation}
w_{\rm eff} < w \; .
\label{wresult}
\end{equation}
In this case, the effect of the vacuum energy density of the
quantized conformally invariant fields is to strengthen the
accelerated expansion that leads to the Big Rip singularity.

If the only conformally invariant quantized fields present were
vector fields, and $ w < w_0$, then $\langle \rho \rangle < 0$,
and, so long as the vacuum energy is considered a small
perturbation ($ \langle \rho \rangle << \rho$), the effect of the
quantized vector fields would be to weaken the evolution towards
the Big Rip singularity, since in this case $w_{\rm eff} > w$.
However, if $ w_0 < w < -1$, then the quantized vector fields
strengthen the Big Rip. The effect of quantized conformally
invariant vector fields is thus to ``push'' $w_{\rm eff}$ towards
the value $w_0$, in which case there is still a Big Rip
singularity caused by divergent expansion. Even quantized
conformally invariant vector fields cannot, within the
semiclassical approximation, weaken a Big Rip singularity to the
point of rendering it regular.

The above calculations were done within the context of the Big Rip
cosmological models constructed by Caldwell et al.
\cite{Caldwell2}, which assume that the parameter $w$ in the
dark-energy equation of state, $w = p/\rho$, remains constant near
the Big Rip, and in which the scale factor, $a(t)$, diverges at a
finite cosmic time. In contrast, the models constructed by Barrow
\cite{Barrow}, in which $w \sim (T_{rip} - t)^{-1-\alpha}$ (with
$\alpha = - 1 + 0^-$) possess a milder form of expansion
singularity, in which the scale factor remains finite (although
its second derivative diverges, with a related divergence in the
pressure) at the singularity. The effects of quantized conformally
invariant fields in such models have been examined by Nojiri and
Odinstov \cite{Nojiri}, who have found that, in those cases, the
quantum effects cause the singularity to become milder.

\acknowledgments
The work of WAH was supported in part by National
Science Foundation Grant No. PHY-0098787.


\end{document}